\documentclass[prl,reprint,twocolumn,superscriptaddress,floatfix]{revtex4}
\usepackage{epsfig}
\usepackage{amsmath}
\usepackage{amssymb}
\usepackage{amsfonts}
\usepackage{mathptmx}
\usepackage{dcolumn}
\usepackage{eucal}
\usepackage{bm}
\usepackage{color}
\usepackage[colorlinks,linkcolor=blue,citecolor=blue]{hyperref}

\usepackage{epstopdf}
\usepackage{graphicx}

\newcommand{\Reff}{R_{\rm eff}}

\begin{document}
\title{A Josephson radiation comb generator}
\author{P. Solinas}
\email{paolo.solinas@spin.cnr.it}
\affiliation{SPIN-CNR, Via Dodecaneso 33, 16146 Genova, Italy}
\author{S. Gasparinetti}
\affiliation{Department of Physics, ETH Z\"urich, CH-8093 Z\"urich, Switzerland}
\affiliation{Low Temperature Laboratory (OVLL), Aalto University School of Science, P.O. Box 13500, 00076 Aalto, Finland}
\author{D. Golubev}
\affiliation{Low Temperature Laboratory (OVLL), Aalto University School of Science, P.O. Box 13500, 00076 Aalto, Finland}
\affiliation{Institute of Nanotechnology, Karlsruhe Institute of Technology, D-76021 Karlsruhe, Germany}
\author{F. Giazotto}
\email{giazotto@sns.it}
\affiliation{NEST, Instituto Nanoscienze-CNR and Scuola Normale Superiore, I-56127 Pisa, Italy}


\maketitle

\textbf{
We propose the implementation of a Josephson Radiation Comb Generator (JRCG) based on a dc superconducting quantum interference device (SQUID) driven by an external magnetic field.
When the magnetic flux crosses a diffraction node of the critical current interference pattern, the superconducting phase undergoes a jump of $\pi$ and a voltage pulse is generated at the extremes of the SQUID. 
Under periodic drive this allows one to generate a sequence of sharp, evenly spaced voltage pulses.
In the frequency domain, this corresponds to a comb-like structure similar to the one exploited in optics and metrology.
With this device it is possible to generate up to several hundreds of harmonics of the driving frequency.
For example, a chain of $50$ identical high-critical-temperature SQUIDs driven at 1 GHz can deliver up to a $0.5$ nW at 200 GHz.
The availability of a fully solid-state radiation comb generator such as the JRCG, easily integrable on chip, may pave the way to a number of technological applications, from metrology to sub-millimeter wave generation.
}

\begin{figure}[t!]
\includegraphics[width=\columnwidth]{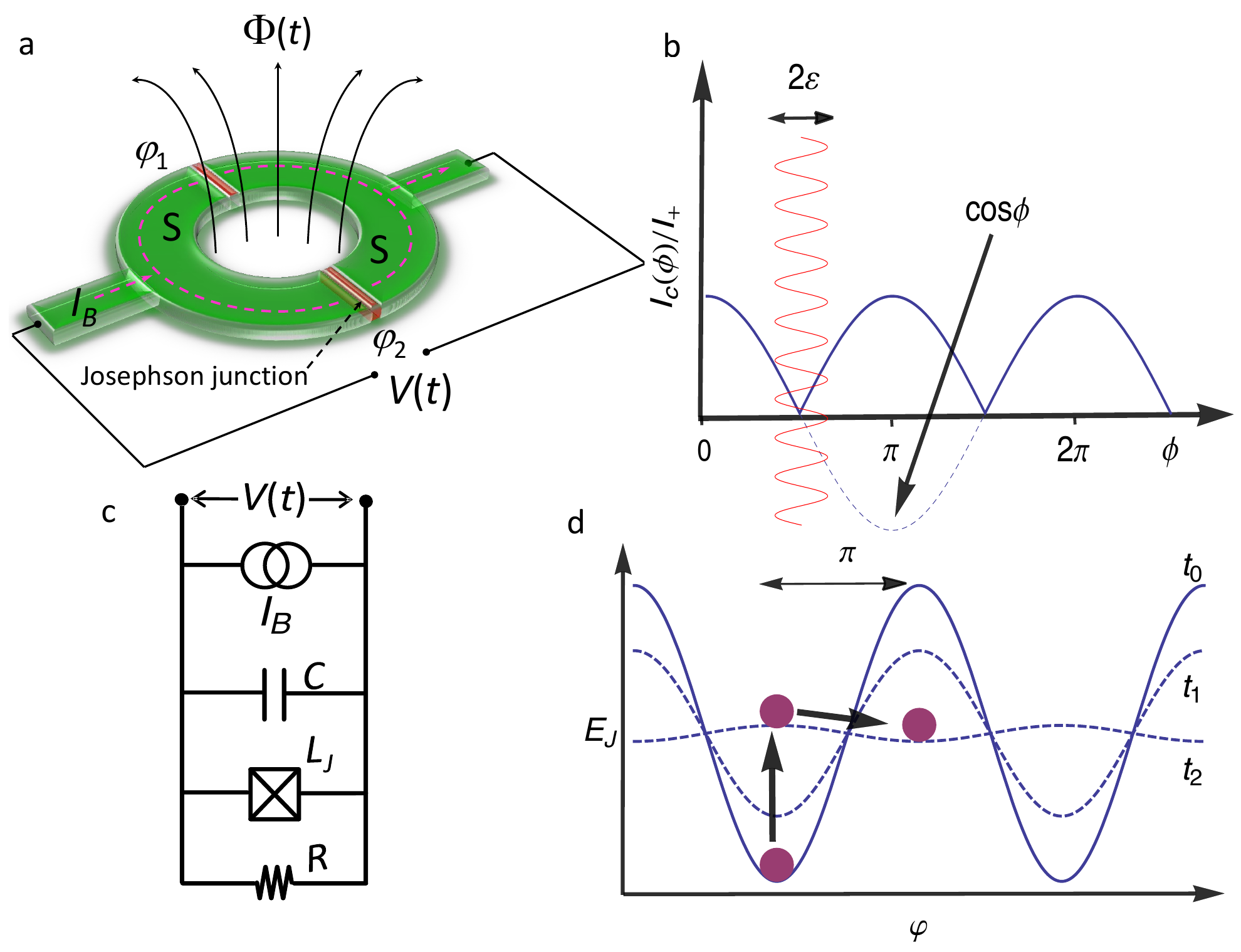}
\vspace{-6mm}
\caption{{\bf  The Josephson radiation comb generator.}
a) A current-biased, flux driven SQUID generates a time-dependent voltage $V(t)$.
The SQUID consists of two superconducting electrodes $S$ (green) connected by two Josephson junctions (red). $\varphi_i$ is the phase across the $i$-th junction, $I_B$ is the constant current bias and $\Phi(t)$ is the time-dependent magnetic flux.
b) Normalized critical current $I_c(\phi)/I_+$ versus normalized magnetic flux $\phi=\pi\Phi/\Phi_0$ for a symmetric SQUID (solid line). The $\phi$-dependent term $\cos \phi$ is also plotted as a dashed line. The phase $\varphi$ across the SQUID undergoes a $\pi$ jump whenever $\phi$ crosses an interference node, due to the change in sign of $\cos \phi$.
Phase jumps can be induced by modulating the flux in time around an interference node with a small amplitude $\epsilon$ (red line).
c) RCSJ model circuit for the SQUID, with resistance $R$, Josephson inductance $L_J$, and capacitance $C$. 
d) Time-dependent tilted-washboard potential for the RCSJ model. The potential is plotted at the initial time ($t_0=0$), at an intermediate time ($t_1= 0.17/\nu$, where $\nu$ is the frequency of the modulation), and just after the vanishing of the potential barrier ($t_2= 0.26/\nu$).
The phase particle (purple ball) starts in an energetic minimum at $\varphi=2 k \pi$. At times later than $1/(4 \nu)$, the position of the particle becomes unstable, leading to a phase jump to the nearest minimum at $\varphi=(2 k+1) \pi$. The direction of the jump is determined by the washboard tilt $\delta=I_B/I_+$, with $\delta \ll 1$.
 }
\label{fig1}
\end{figure}

Optical frequency combs have been a major research trend of the last decade \cite{udem2002optical}.
The possibility to generate higher harmonics starting from a fundamental one has made it possible to extend the accuracy of the atomic clocks from the radio to the optical frequency region, leading to breakthroughs in optical metrology \cite{hansch1999laser}, high precision spectroscopy \cite{bloembergen1977nonlinear, hansch1994frontiers} and telecommunication technologies \cite{udem2002optical, foreman2007remote}.
Here we show that a similar-in-spirit harmonic generator can be implemented with a dc superconducting quantum interference device (SQUID) subject to a time-dependent magnetic field.
Driven by the field, the superconducting phase difference across the SQUID undergoes jumps of $\pi$, which are associated to a sequence of sharp voltage pulses. This pulse sequence translates into a radiation comb in frequency domain, thereby realizing a Josephson radiation comb generator.
This device could have applications extending from the precision frequency measurement (as in the optical analogue) to the use as a sub-millimiter wave generation.
The main advantages are the possibility to fabricate it on-chip and its integrability within the standard electrical circuits.

Our proposal for a JRCG is based on a dc SQUID (see Fig. \ref{fig1}a), consisting of two Josephson junctions arranged in parallel in a superconducting loop. The SQUID is biased by a constant current $I_B$ and it is driven by an external, time-dependent magnetic flux $\Phi$. Here we assume the inductance of the loop to be negligible with respect to the Josephson inductance of the junctions. 
Due to the first Josephson relation \cite{tinkham2012introduction}, the current ($I_J$) vs phase relation of the SQUID reads 
\begin{equation}
I_J(\varphi; \phi) = I_+ [\cos \phi \sin \varphi + r~ \sin \phi \cos \varphi], \,
\label{eq:I_J}
\end{equation}
where $\varphi=(\varphi_1+\varphi_2)/2$,  $\phi = \pi \Phi/\Phi_0$ ($\Phi_0\simeq 2 \times 10^{-15}$ Wb is the flux quantum), $I_+ = I_{c1} + I_{c2}$, $\varphi_i$ and $I_{ci}$ ($i=1,2$) are the phase across and the critical current of the $i$-th junction, respectively,
and  $r = (I_{c1} - I_{c2})/ (I_{c1}+I_{c2})$ expresses the degree of asymmetry of the interferometer.
Equation (\ref{eq:I_J}) describes the well-known oscillations of the SQUID critical current $I_c(\phi) = \max_\varphi I_J(\varphi;\phi)$ as a function of the magnetic flux, with minima occurring at integer multiples of $\Phi_0/2$ (see Fig. \ref{fig1}b) \cite{tinkham2012introduction}, and it already contains the main feature of the effect we want to discuss.
Let us consider the behavior of the phase $\varphi$ as $\Phi$ crosses a critical-current minimum and take a symmetric SQUID ($r=0$) for simplicity.
If the biasing current is fixed, then we see from Eq.~(\ref{eq:I_J}) that a change of sign in $\cos \phi$ must be accompanied by a change of sign in $\sin \varphi$ in order for the current to maintain its direction. This change of sign is accomplished by a phase jump of $\pi$ \cite{giazotto2012josephson, giazotto2013coherent, martinez2014quantum}, which, owing to the second Josephson relation \cite{tinkham2012introduction}, results in a voltage pulse $V(t)$ across the SQUID.

\begin{figure}[t!]
\includegraphics[width=\columnwidth]{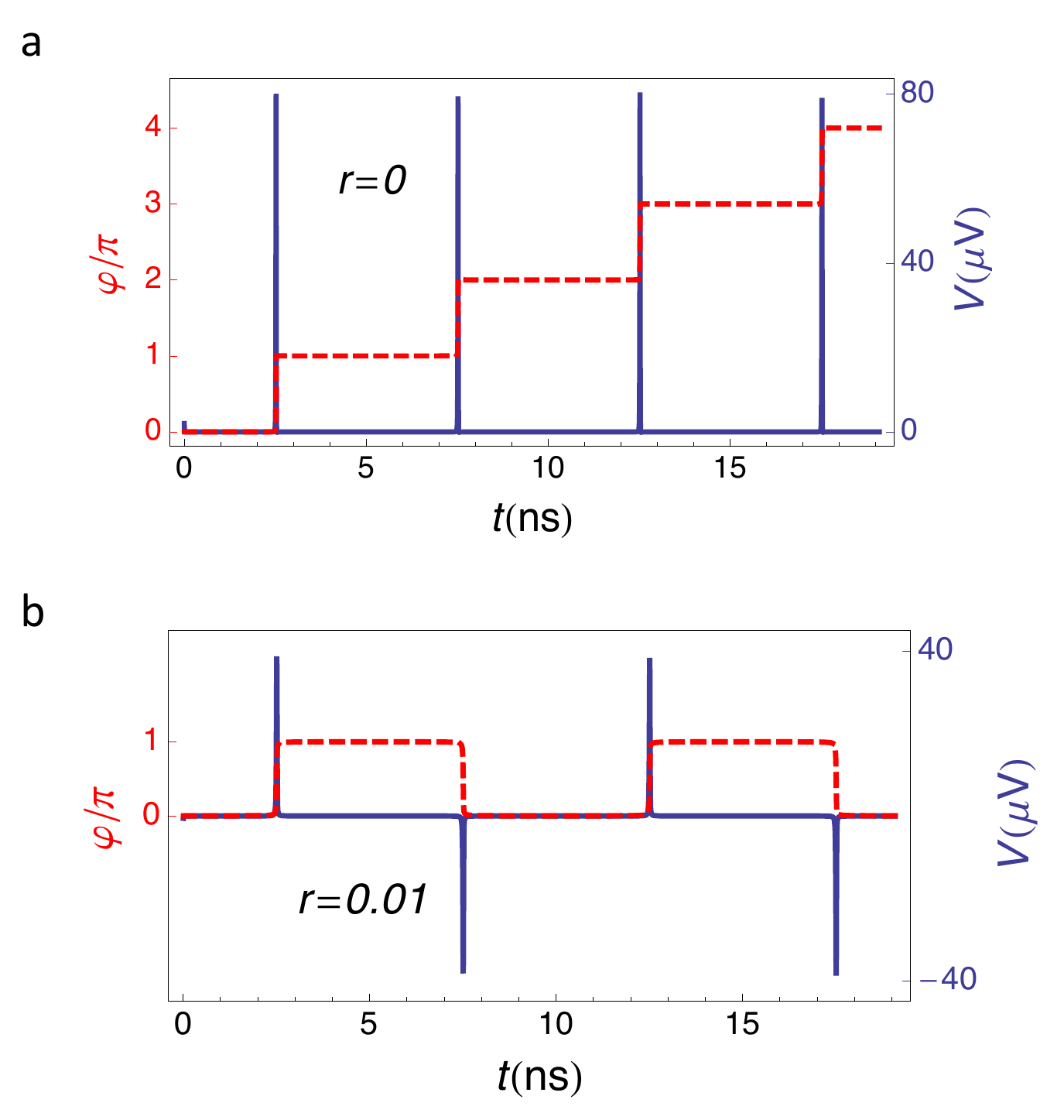}
\vspace{-6mm}
\caption{{\bf Phase jumps and voltage comb.} 
Time evolution of the phase $\varphi$ (dashed line, left axis) and corresponding voltage $V$ developed across the SQUID (solid line, right axis).
Voltage pulses are generated at times $(2 k+1)/4\nu$ (with $k$ integer), when an interference node is crossed. The evolution is shown for a symmetric SQUID (a) with $r=0$ and $I_B= 10^{-3} I_+$, and for an asymmetric SQUID (b) with $r=0.01$ and $I_B= 0$.
In (a), the current bias determines the direction of the phase jumps.
In (b), the SQUID asymmetry induces alternate phase jumps even in the absence of an external current bias. 
The driving frequency is $\nu=100~$MHz and the amplitude is $\epsilon=0.9$.
The SQUID parameters are typical for a Nb/AlOx/Nb junction \cite{patel1999self}, with $R=10~$ Ohm, $I_+ = 0.2$ mA and $I_+ R = 0.2~$mV. The corresponding capacitance would be of the order of $=10~$fF and has been neglected. 
}
\label{fig2}
\end{figure}

For a quantitative characterization of the phase jumps, we need to study the dynamics of the phase. To do so, we rely on the resistively and capacitively shunted Josephson junction (RCSJ) model \cite{tinkham2012introduction, gross2005applied}. We model the SQUID as a capacitor $C$, a resistor $R$, and a non-linear, flux-dependent inductor $L_J$ arranged in a parallel configuration (see Fig.~\ref{fig1}c). 
We consider a sinusoidally-driven magnetic flux with frequency $\nu$ and amplitude $\epsilon$, centered in the first node of the interference pattern, so that $\Phi(t) = \Phi _0/2[1-\epsilon  \cos (2 \pi \nu t )]$. As a result, the magnetic flux crosses the nodes of the interference pattern at $t= (2k+1)/ 4 \nu$, with $k$ integer.
The equation for $\varphi$ can be written in terms of the dimensionless variable $\tau = 2 \pi \nu t$ as \cite{gross2005applied}
\begin{equation}
   c \frac{d^2 \varphi}{d \tau^2}+  \frac{d \varphi}{d \tau} + \alpha[ f(\varphi,\tau) - \delta]=0,
  \label{eq:RCSJ_adim}
\end{equation}
where $ \delta=I_B/I_+$, $c =2 \pi R C \nu$, $f(\varphi,\tau) = I_J[\varphi;\phi(\tau)] / I_+$ and $\alpha= I_+ R/(\Phi_0 \nu)$.

Equation \ref{eq:RCSJ_adim} is usually interpreted in terms of a fictitious phase particle moving in a tilted-washboard Josephson potential $E_J$, as shown in Fig.~\ref{fig1}d \cite{tinkham2012introduction}. Here we restrict ourselves to small biasing current ($\delta \ll 1$), corresponding to a small tilt. Furthermore, we focus on the limits $c \ll 1$ (overdamped regime) and $|\alpha| \gg 1$, as these two conditions maximize the JRCG performance (see SI).

We first consider a symmetric SQUID ($r=0$). 
Then the time-dependent Josephson potential is $E_J (t) =\int{I_{tot}V(t)dt}= - E_{J0}[ f(t)\cos \varphi + \delta \varphi]$ where $f(t) = \cos(\pi \Phi/\Phi_0)$, $E_{J0}=\Phi_0^2\nu\alpha/(2\pi R)$, and $I_{tot}=I_J-I_B$ \cite{tinkham2012introduction,gross2005applied}.
When $t=0$ the potential has minima at $\varphi = 2 k \pi$.
For $t= 1/(4 \nu)$ the potential barrier vanishes and $E_J = -E_{J0} \delta \varphi$. For $t> 1/(4 \nu)$, $f(t)$ changes sign and the potential minima occur at $\varphi = (2 k +1) \pi$.
The former equilibrium points $\varphi=2k\pi$ have become unstable and the system tends to move to one of the new minima, resulting in a $\pi$-jump in the phase. 
This cartoon picture helps us to pinpoint the difference between the phase jumps discussed in this work, the $2 \pi$-phase slips appearing in low-dimensional superconductors 
\cite{arutyunov2008superconductivity,astafiev2012coherent,langer1967intrinsic,zaikin1997quantum} and the 
$2 \pi$-phase jumps used in the rapid single flux quantum (RSFQ) logic \cite{likharev1985resistive, mukhanov1987ultimate}.
$2\pi$-phase slips typically stem from thermal activation or quantum fluctuations.
As for the RSFQ $2\pi$-phase jumps, they are generated by a current pulse in an otherwise static potential landscape.
By contrast, in the JRCG the magnitude of the jumps is $\pi$ and the jumps have a purely energetic origin.

\begin{figure}[t!]
\includegraphics[width=\columnwidth]{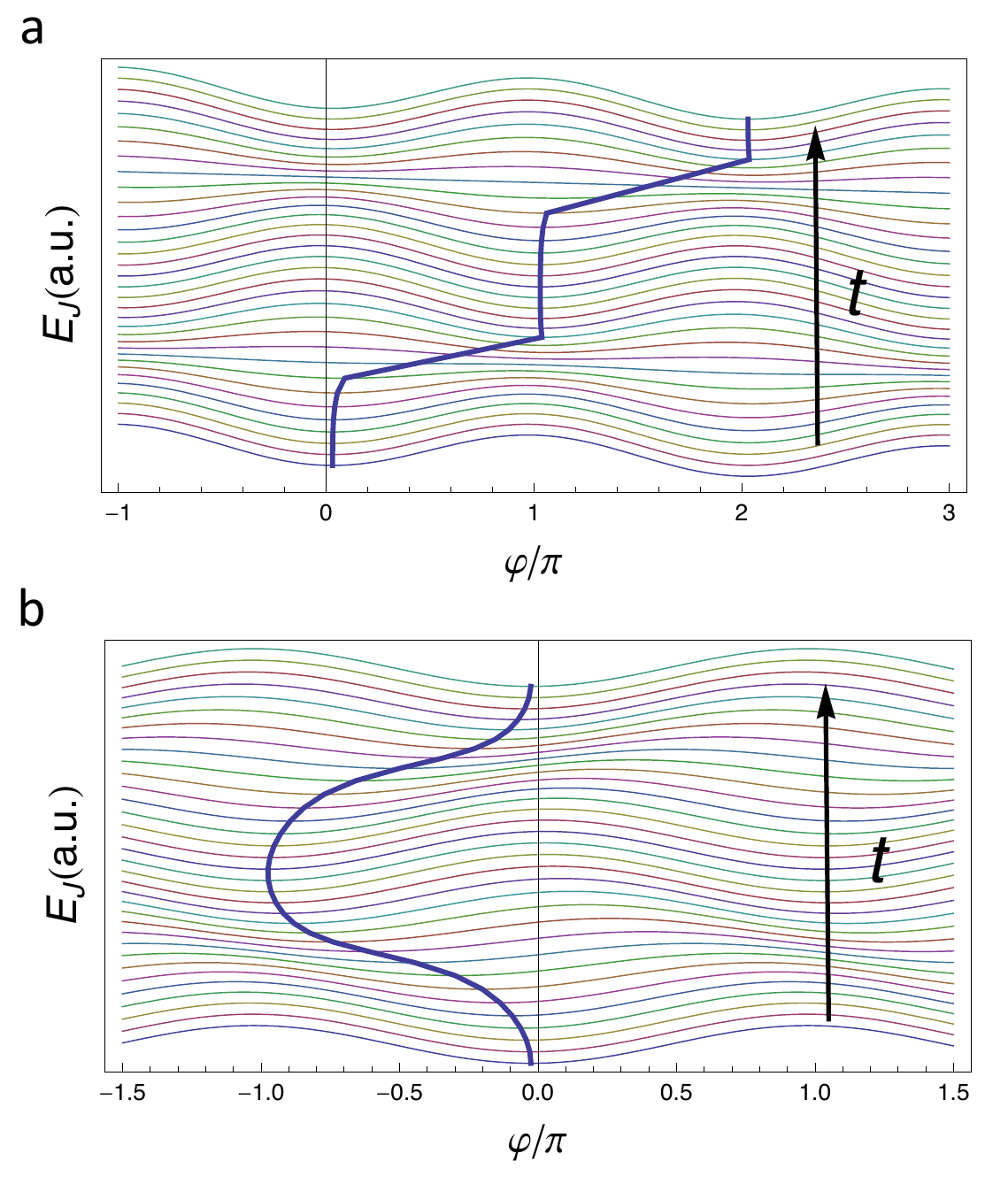}
\vspace{-6mm}
\caption{{\bf  Tilted-washboard potential for symmetric and asymmetric SQUID.}
The potential landscape $E_J(\varphi)$ is plotted at increasing times $t$. The traces are stacked by a constant offset on the vertical axes. A blue, thick curve tracks the position of a local minimum of the potential at different times.
a) Symmetric SQUID, i.e., $r=0$, with tilt $\delta = 0.3$. 
b) Asymmetric SQUID, with $r=0.5$ but $\delta = 0$. 
For presentation purposes, we have chosen larger values of $r$ and $\delta$ than those giving the best performance for the device (see Fig.~\ref{fig4}).
 }
\label{fig3}
\end{figure}

The numerical solution of Eq.~(\ref{eq:RCSJ_adim}) for $r=0$ is shown in Fig.~\ref{fig2}a.
As the critical current crosses the minimum at $\Phi = \Phi_0/2$, the phase experiences a $\pi$-jump and a voltage pulse is generated across the SQUID.
The shape of the pulse is determined by the parameter $I_+ R$ (see SI): the {\it larger} $I_+ R$,  the {\it sharper} the voltage pulse. 
We notice that the presence of a finite bias current $I_B$ is crucial to impose a preferred direction to the phase jumps (see also Fig.~\ref{fig3}a).
The same analysis essentially holds as well for a weakly-asymmetric SQUID ($r \ll 1$), as long as $I_B$ is strong enough to force the phase particle to roll always in the same direction.
However, the junctions asymmetry brings in a key ingredient to the JRCG, which becomes apparent in the limit $I_B\to 0$. Indeed, a finite asymmetry imposes an alternate pattern to the phase jumps (see Fig.~\ref{fig3}b and SI). This realizes an ideal ac pulse source. 

We can explain this behavior following the analogy with the phase particle in a time-dependent potential.
For $r \neq 0$, the position of the minima changes in time (see Fig. \ref{fig3} and SI).
This means that if the system starts in a minimum at $t=0$, it is close but not in a maximum when the time-dependent potential changes sign.
This small deviation from the maximum point induces the phase particle rolling and the corresponding phase jump even in absence of current bias.
For a periodic drive, the particle is found alternatively on the left and on the right of the maximum; as a result, it rolls in alternate directions producing the alternate pattern of the voltage pulses.
In Fig. \ref{fig3}b, we show the potential $E_J$ vs $\varphi$ at different time. 
The blue thick curve represents the position of the minima of the potential.
In the absence of a current bias, the phase undergoes a sequence of positive and negative jumps resulting in the alternate voltage pulses.
This device configuration realizes an ideal ac pulse source. Furthermore, the limit $I_B\to0$ corresponds to a floating device. This facilitates the integration of the JRCG in microwave-based architectures such as circuit-QED \cite{blais2004cavity, wallraff2004strong, koch2007charge}.

\begin{figure}[t!]
\includegraphics[width=\columnwidth]{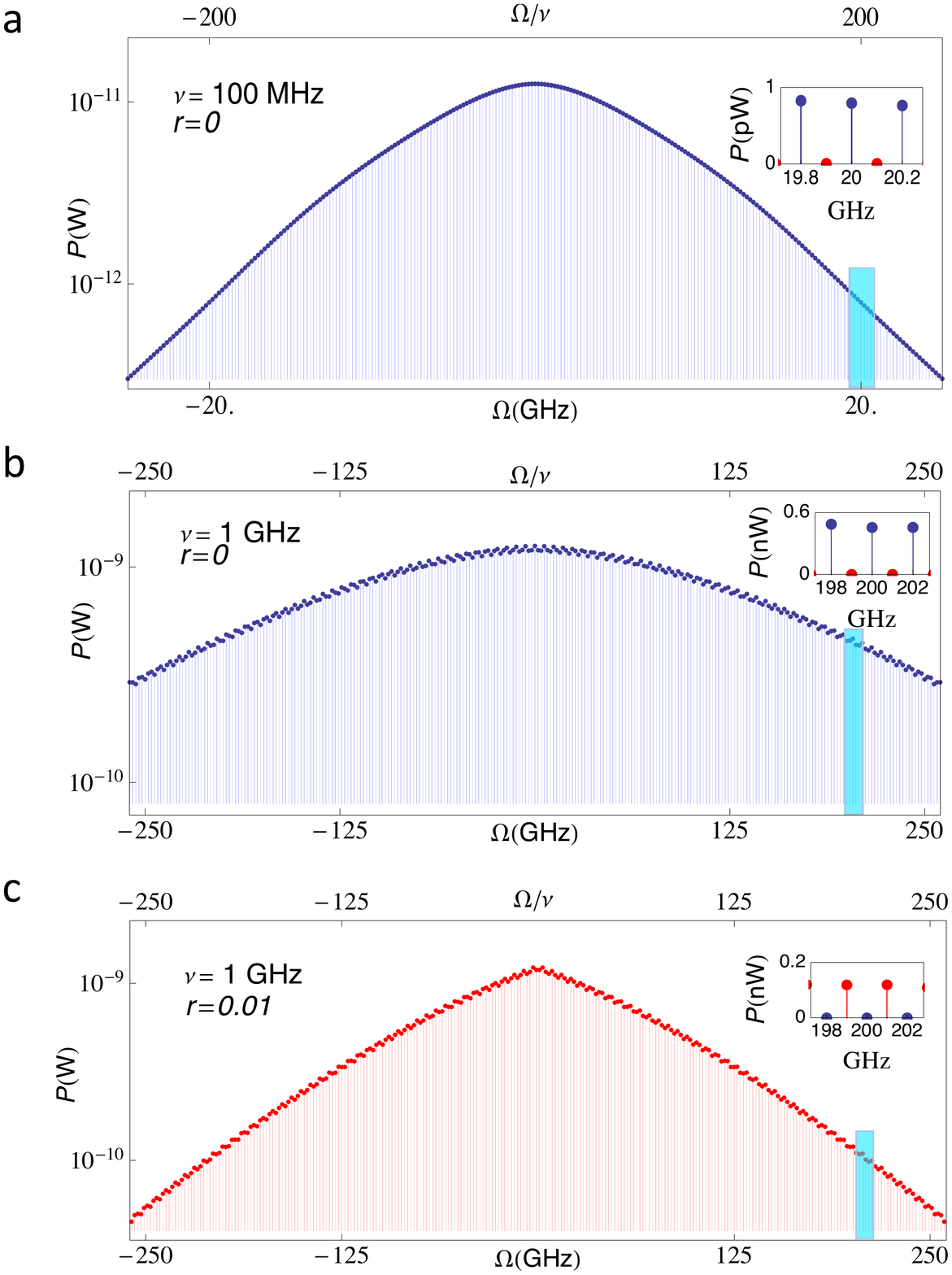}
\vspace{-6mm}
\caption{ {\bf Power spectrum of the Josephson radiation comb generator.}
We consider a chain of $N=50$ SQUIDs coupled to a $50~\Omega$ load.
To emphasize the behavior at high frequency we use a logarithmic scale in the main panels but keep the linear scale in the insets. The cyan regions correspond to the insets. 
(a) Power spectrum $P$ vs frequency $\Omega$ for a symmetric ($r=0$) Nb/AlOx/Nb SQUID chain, driven at $\nu =100~$MHz. The SQUID parameters are the same as in Fig.~\ref{fig2}. Due to the presence of the load, each SQUID sees an effective resistance $R_{\rm eff} \approx 1~$Ohm.
The output power at $20$~GHz ($200$-th harmonic) is about $1~$pW (see Inset).
(b) $P$ vs $\Omega$ for a symmetric ($r=0$), high-critical temperature YBCO SQUID chain, with $\nu = 1~$GHz, $I_+ R = 10$ mV, $\Reff=1~$Ohm and $I_B= 10^{-3}I_+$  \cite{malnou2014high,rosenthal1994}. At $200~$GHz ($200$-th harmonic) $P$ is about $0.5~$nW (see Inset).
(c) $P$ vs $\Omega$ for an asymmetric ($r=0.01$) YBCO SQUID chain, with $\nu = 1~$GHz, $I_+ R = 10$ mV and $I_B= 0$.
In all panels, the drive is such that $\Phi(t)$ oscillates around the first interference node ($\Phi_0/2$) with elongation $\epsilon=0.9$. Blue and red dots are used to plot the even and the odd harmonics, respectively. In (a,b) only the even harmonics are present due to the comb-like shape of the voltage pulses. In (c), due to the alternating direction of the voltage pulses, only the odd harmonics are present. The junction capacitance has been neglected in all calculations as it does not affect the dynamics. 
}
\label{fig4}
\end{figure}

The voltage pulses shown in Fig \ref{fig2} suggest an application similar to the frequency combs used in optics \cite{udem2002optical}. 
In this context, the most relevant feature becomes the sharpness of the voltage pulse, which is related to the number of harmonics generated. 
The sharpness is essentially determined by $I_+ R$, which, in turn, depends on the material properties of the Josephson junctions.
While the total output power  provided by a single JRCG is fairly small, it can be boosted by using an array of nominally-identical SQUIDs. A similar approach is used for the realization of the metrological standard for voltage based on the Josephson effect \cite{shapiro1963josephson, kautz1987precision,tsai1983high, gross2005applied,lloyd1987,popel19901, cybart2014large}. Before presenting our results for an array of SQUIDs, we discuss the approximations we have used in our analysis.

First of all, we have neglected the coupling between the SQUIDs via mutual inductance and/or cross capacitance and inductance of the superconducting wire. This condition, which can be realized in practice by a suitable design choice, implies that the dynamics of each SQUID is independent from the rest of the chain. As a matter of fact, due to current conservation, the current $I_i$ through the $i$-th SQUID is the same and equal to $I_B$: $I_i=I_B$. 
For every SQUID we can write the total current as $I_i = I_{i,S} + I_{i,N} + I_{i,C}= I_B$ \cite{gross2005applied} where $I_{i,S}= I_{i,C}  \cos(\pi \Phi/\Phi_0)\sin \varphi_i$ is the superconducting current (for $r=0$), $I_{i,N} = V_i/R_i$ is the normal current and $I_{i,D}= C_i dV_i/dt$ is the displacement current. In the above expressions, $V_i$ and $\varphi_i$ are the voltage and the phase across the $i$-th SQUID, $C_i$ and $R_i$ are the  capacitance and the resistance of the SQUID, respectively.
By using the Josephson voltage-phase relation we then obtain the RCSJ Eq.~(\ref{eq:RCSJ_adim}) for the $i$-th SQUID, which is independent from the other SQUIDs in the array.
Given this result, it is possible to obtain the voltage at the extremes of the array by summing up the voltage of the single SQUIDs: $V(t)= \sum_{i} V_i(t)$.

So far we have discussed the voltage produced by the JRCG in the absence of any external load.
From an experimental point of view, a quantity of greater interest is the power that can be transferred to a given load $R_L$. The effect of a finite load on the chain can be understood in terms of an additional current flowing through the chain. Under the assumption that all the SQUIDs are identical, the problem can still be treated exactly. It turns out (see SI) that the voltage across the load is still given by $\sum_{i} V_i(t)$, provided an additional shunt resistor $R_L/N$ is added in parallel to each SQUID. The shunt resistance $R$ in the RCSJ equation (\ref{eq:RCSJ_adim}) must then be replaced by an effective resistance $\Reff = R R_L/ (R_L+N R)$. We find that the delivered power $P=V(t)^2/R_L$ scales as $N^2$ as long as $N \ll R_L/R$, while in the opposite limit $N \gg R_L/R$ it scales as $N$.


A further assumption we have made is that the emitted radiation propagates instantly across the device. Such lumped-element model is certainly appropriate for a short SQUID chain but will eventually break down as the total length of the chain approaches the wavelength of the emitted radiation. In that limit, the chain must be regarded as a distributed element and we generally expect the frequency comb to be distorted by wave-interference effects. 
The relation between the minimum wavelength $\lambda_{\rm min}$ and the chain length $L$ relation for the validity of the lumped-element model is (see SI): $\lambda_{\rm min} \geq L/2$. However, even when this condition is no longer satisfied, a suitable choice of the effective distance between the SQUIDs can ensure constructive interference at a specific frequency. This feature can be exploited to operate the device at higher frequencies and/or with higher output.

At this point, we can present the predicted performance of the device. In Figure \ref{fig4} we show the calculated JRCG power spectrum $P$ vs frequency $\Omega$ (see SI) for two driving frequencies and for different junctions and symmetry parameters. 
Figure \ref{fig4}a) displays the behavior of a chain of $N=50$ symmetric Nb/AlOx/Nb SQUIDs \cite{patel1999self} with a $100~$MHz drive. The parameters are the same as those in Fig.~\ref{fig2}a. 
The sharp pulses determine the broad range of the emitted radiation, up to several hundreds of harmonics. 
At $20~$GHz (see the inset of Fig.~\ref{fig4}a) the JRCG provides an output power of $\sim 1\,$pW. This power level can be detected, for instance, by coupling the device to a transmission line and feeding the signal to a commercial spectrum analyzer.

In order to achieve sharp pulses at higher frequencies, one needs to use a superconductor with a larger characteristic voltage ($I_+R$). In such a way, one can drive the SQUID at higher frequencies.
In Fig.~\ref{fig4}b we show the results expected for a symmetric YBCO SQUID series \cite{rosenthal1994,malnou2014high, cybart2014large} at $1~$GHz drive.
YBCO Josephson junctions provide a large superconducting gap with a characteristic voltage $I_+ R \approx 10$ mV and possess a negligible intrinsic capacitance \cite{rosenthal1994}.
Due to the larger driving frequency, the emitted signal at $200~$GHz is still sizable, reaching an output power of a fraction of nW (Fig.~\ref{fig4}b, inset).
Such a signal is already in the far infrared range, which has seen a substantial research development in the last two decades due to countless technological applications.
In this frequency range, the radiated signal can be coupled to free space through, for instance, an antenna coupled to the SQUID electrodes \cite{vicarelli2012graphene,erickson2014frequency}.
The power spectrum is similar for an asymmetric YBCO SQUID chain (Fig.~\ref{fig4}c).
The main differences lie in the presence of odd harmonics only and in a smaller output power at high frequency (around $0.2$nW at 200 GHz).

The device has room for optimization.
The delivered power at high frequency can be increased by suitable array design, i.e., either by using different materials or optimizing the inter-SQUID distance, or by using parallel configuration of JRCGs (see SI).



We have analyzed the effect of thermal noise at 4.2 K on the device performance (see SI).
Noise is expected to be the most harmful in the vicinity of the phase jumps, as the potential barrier is the most shallow there (see Fig.~\ref{fig1}d).
However, our numerical calculations show that its effect is negligible for the parameters used in Fig.~\ref{fig4}. 
Thermal noise could play a role at slow driving frequencies because it is easier to induce undesired transitions when the the potential barrier is shallow. However, this effect can be counteracted by increasing the current $I_B$ to impose a privileged direction to the dynamics (see SI).
In the array configuration, particular care must also be taken to avoiding random flux offsets, due to, for instance, trapped vortices. Such offsets would cause the SQUIDs to switch at different times and thereby contribute to the smearing of the voltage comb features. Similar detrimental effects can be produced by imprecision in the SQUID fabrication, i.e., in the asymmetry parameter and SQUID area. 

The proposed JRCG is within the reach of state-of-the-art nanofabrication technology. SQUID arrays with an asymmetry dispersion of the order of $\sim 0.05-1\%$ can be fabricated with standard lithographic techniques. Furthermore, a single on-chip superconducting line can be used to drive the magnetic fluxes of a SQUID array in a synchronized manner and with ns time resolution. The delivered power at high frequency may be increased beyond our estimates by fabrication and/or design, for instance, by using different materials or by operating more JRCGs in a parallel configuration (see SI).


In summary, we have proposed a Josephson radiation comb generator based on a flux-driven SQUID array. Based on our preliminary analysis, its implementation seems realistic and may pave the way to a number of applications, from low-temperature microwave electronics to on-chip sub-millimiter wave generation.


We gratefully acknowledge R. Bosisio, A. Braggio, M. Hofheinz, M. J. Mart\'inez-P\'erez, M. Pechal, and A. Tredicucci for fruitful discussions. 
P.S. has received funding from the European Union FP7/2007-2013 under REA
grant agreement no 630925 -- COHEAT and from MIUR-FIRB2013 -- Project Coca (Grant
No.~RBFR1379UX).
The work of F.G. was partially supported by the Marie Curie Initial Training Action (ITN) Q-NET 264034, and by the European Research Council under the European Union's Seventh Framework Program (FP7/2007-2013)/ERC grant agreement No. 615187-COMANCHE.
S.G. acknowledges financial support from the Aalto University network in Condensed Matter and Materials Physics (CMMP) and from the Swiss National Science Foundation (SNF) Project 150046.



\newpage

\clearpage
\setcounter{equation}{0}
\setcounter{figure}{0}
\setcounter{section}{0}

\section{\Large Supplementary Material}

\section{Solution of the RCSJ equation in a dc SQUID}
\label{sec:RCSJ_eq}

We consider a dc SQUID composed by two Josephson junctions and subject to a magnetic flux $\Phi$.
The total current though the SQUID is $I_J= I_{c1}  \sin \varphi_1 + I_{c2}  \sin \varphi_2$, where $I_{ci}$ and $\varphi_i$ are the critical current and the phase across the $i$-th junction, respectively.
Because of the flux quantization constraint, it follows that $(\varphi_1-\varphi_2)/2 = \pi \Phi/\Phi_0$. Introducing the phase across the SQUID $\varphi= (\varphi_1+\varphi_2)/2$, we get 
\begin{equation}
I_J[\varphi;\phi(\tau)] = I_+ [\cos \phi \sin \varphi + r~ \sin \phi \cos \varphi] \,,
\label{eq:current}
\end{equation}
where $\phi = \pi \Phi/\Phi_0$, $I_+ = I_{c1} + I_{c2}$, $r = (I_{c1} - I_{c2})/ (I_{c1}+I_{c2})$ and  $\Phi_0\simeq 2\times 10^{-15}$ Wb is  the flux quantum. 

Starting from the RCSJ model \cite{tinkham2012introduction, gross2005applied}, we can write an equation of motion for the phase $\varphi$ as
\begin{equation}
  \frac{\hbar C}{2 e }  \ddot{{\varphi}} + \frac{\hbar}{2 e R} \dot{\varphi} + I_+  f(\varphi,t) = I_B
  \label{eq_app:RCSJ}
\end{equation}
where $C$ is the capacitance, $R$ is the total shunting resistance of the SQUID, $I_B$ is the external biasing current and $f(\varphi,t) = I_J[\varphi;\phi(t)] / I_+$.
We rescale the above equation in terms of the driving frequency $\nu$: $\tau = 2 \pi \nu t$. Using $\hbar/(2 e)= \Phi_0/(2 \pi)$, we obtain
\begin{equation}
 c \frac{d^2 \varphi}{d \tau^2}+  \frac{d \varphi}{d \tau} + \alpha[ f(\varphi,\tau) - \delta]=0,
  \label{eq_app:RCSJ_adim}
\end{equation}
where $ \delta=I_B/I_+$, $c = 2 \pi R C \nu$ and 
\begin{equation}
 \alpha= \frac{I_+ R}{\Phi_0 \nu}.
 \label{eq_app:alpha}
\end{equation}

\subsection{Analytical solution for $I_B=0$}
\label{app:no_bias_solution}

Let us consider the case of a symmetric dc SQUID ($r=0$), overdamped junctions ($c\approx 0$) and zero current bias ($\delta=0$). Then (\ref{eq_app:RCSJ_adim}) reduces to
\begin{equation}
     \frac{d \varphi}{d \tau} + \alpha F(\tau) \sin \varphi =0,
  \label{eq:RCSJ_overdamped}
\end{equation}
where $F(\tau) = \cos [\phi(\tau)]$.
We notice that if the initial condition is $\varphi(0) = \varphi_0 = k \pi$, the above equation has trivial dynamics $\varphi(t)=0$. This means that, even if small, we cannot neglect the influence of the $\ddot{{\varphi}}$ term.  
However, if $\varphi_ \neq k \pi$, we can effectively neglect the capacitive contribution and solve analytically the differential equation (\ref{eq_app:RCSJ_adim}) to obtain 
\begin{equation}
 \varphi(t) = 2 \arctan \Big[ \exp{\Big(-\alpha \int_0^t d \tau F(\tau) \Big)}  \tan \frac{\varphi_0}{2}\Big].
 \label{eq:varphi_sol}
\end{equation}

We suppose that  $|\alpha| \gg 1$ (and $\alpha>0$).
If $\int_0^t d \tau F(\tau)$ assumes positive and negative values, the argument of the arctangent increases or decreases exponentially depending on its sign.
For simplicity, we consider $\int_0^t d \tau F(\tau)<0$ and the case of small $\varphi_0$.
For $\varphi_0>0$, $\varphi(t)$ exponentially reaches $\pi$, and for $\varphi_0<0$, the evolution is similar but $\varphi(t)$ varies between $\varphi_0$ and $-\pi$.
Therefore, in both cases we have an exponential $\pi$ jump of the phase but its direction is determined by the initial phase $\varphi_0$.
The rate of the exponential jump is determined by $I_+ R$: the larger $I_+ R$, the sharper the voltage pulse.

Recalling the phase particle analogy discussed in the main text, with $c=0$ if the phase is initially in the minimum 
$\varphi_0 = 2 k \pi$ it will remains in the same point even when it becomes an unstable maximum. 
A small shift of the initial condition induces an exponential dynamics since at $t= 1/(4 \nu)$ the phase is close to (but not on) a potential maximum.
The direction of this shift (and, in the case discussed, the sign of $\varphi_0$) determines the direction of the fall and the jump of the phase.

The change in time of $\varphi$ in Eq. (\ref{eq:varphi_sol}) is associated to a voltage
\begin{eqnarray}
  \frac{2 e }{\hbar}V(t)&=& \frac{- 2 \pi \nu \alpha ~ F(t) \sin \varphi_0 }{\sin ^2\left(\frac{\varphi_0 }{2}\right)
   e^{-\alpha  \int_0^t f(\tau ) \, d\tau }+\cos ^2\left(\frac{\varphi_0
   }{2}\right) e^{\alpha  \int_0^t f(\tau ) \, d\tau }}   \label{eq:V_t}
   \label{eq:V}
   \end{eqnarray} 
(Notice that we have gone back the real time $t=\tau/(2 \pi \nu)$ unit.)

For a small oscillation around the first diffraction node it is possible to have an analytical expression for $V(t)$.
In this case, $\cos(\pi \Phi/\Phi_0) \approx  (\pi/2)  \epsilon  \cos (2 \pi  \nu  t)$.
After a straightforward calculation, we approximate the function for small times, i.e., $t\ll 1/\nu$ to obtain
\begin{equation}
  V(t) = -\frac{\pi}{2}   \alpha  \nu  \Phi _0 \epsilon \cosh^{-1} \left( \frac{\pi  \alpha   \epsilon t \nu}{2}  + \rho_0 \right)
\end{equation}
where $\rho_0 = \log \left[ \tan( \varphi_0/2) \right]$.
The above expression has an exponential behavior (increasing and decreasing) when the system crosses a diffraction node.

The mean value of the voltage square, which is related to the delivered power, is given by $ \langle V^2(T) \rangle  = (1/T) \int_0^T dt V^2(t)$, where $T$ is the averaging time. 
Taking into account Eq. (\ref{eq_app:alpha}) and for long average time $T \gg 2/(\pi  \alpha  \epsilon \nu)$, we obtain
\begin{equation}
 \langle V^2(T) \rangle =\frac{ \Phi _0 \epsilon I_+ R}{4 T} \tanh \left( \frac{\pi  \alpha   \epsilon T \nu}{2}  + \rho_0 \right) \approx \frac{ \Phi _0 \epsilon I_+ R}{4 T}.
 \label{eq:app_V^2}
\end{equation}

\subsection{Solution in presence of a current bias}

We consider now the case of a small biasing current $I_B$ in a symmetric SQUID ($r=0$).
By "small biasing current", we mean  that its effect must be negligible with respect to the driven dynamics, i.e., $\delta \ll \alpha$, but must dominate the capacitor dynamics, i.e., $\delta \gg c$.

When $\varphi \approx k \pi$, the Josephson current contribution $\alpha F(\tau) \sin \varphi $ is small and the dynamics is determined only by $I_B$.
Equation (\ref{eq_app:RCSJ_adim}) reduces to 
\begin{equation}
 \frac{d \varphi}{d \tau} - \alpha   \delta =0.
 \label{eq:eq1_const_I}
\end{equation}
Away from $\varphi \approx k \pi$, the Josephson current contribution dominates and, therefore, the equation for motion approximatively reads
\begin{equation}
 \frac{d \varphi}{d \tau} + \alpha F(\tau) \sin \varphi(\tau)=0.
 \label{eq:eq2_const_I}
\end{equation}
Since the behavior of $\varphi(t)$ under these two different dynamics is drastically different (linear versus exponential change), we can tune the system parameters in order to effectively separate the two regimes governed by Eqs. (\ref{eq:eq1_const_I}) and (\ref{eq:eq2_const_I}) and generate the sequence of exponential phase jumps.

In other terms, the presence of a current bias has two effects.
The first is to transport the system away from the region $\varphi \approx k \pi$ in which the drive contribution is small and the dynamics is dominated by the capacitive term. 
The second is to breaks the symmetry of the system inducing the jumps always in the same direction.

The effect of a current bias can be interpreted in terms of the tilted-washboard potential discussed in the main text. 

\subsection{Asymmetric SQUID case}

From Eq. (\ref{eq:current}), the energy potential reads (we neglect the current bias and set $\delta=0$) \cite{tinkham2012introduction}
\begin{equation}
 E_J(t) =  -\cos \phi \cos \varphi + r\sin \phi \sin \varphi.
\end{equation}
To find the position of the maxima and the minima we derive $E_J(t)$ with respect to $\varphi$ and equal it to zero.
The corresponding equation reads $\tan \varphi  = r \tan  \phi $.
For $r=0$, we see that the values of $\varphi$ satisfying the above equation do not depend on time.
On the contrary, since $\phi = \pi \Phi/\Phi_0$, they depend on time through $\Phi$ for any $r \neq 0$ as discussed in the main text.
For a periodic drive, the particle is found alternatively on the left and on the right of the maximum and it rolls in alternate directions producing the alternate pattern of the voltage pulses.



\section{Finite-size effects}

We consider a long chain of SQUIDs located at $x = x_k$, ($k = 1,2,...,N$) and coupled to a transmission line of
length $L$. We assume that the first SQUID is at point $x_0 = 0$ and that $x_k = ka$, where $a$ is the distance between the SQUIDs. Every SQUID emits radiation with the spectrum $v(\omega)$. The time dependent voltage drop across it is $V(t) = \int d\omega/ (2 \pi) \exp{(- i \omega t) v(\omega)}$. In case of periodic drive one should have $v(\omega)=\sum_n V_n \delta(\omega - n \omega_0)$, where $\omega_0$ is the driving frequency. Then the voltage at the end of the transmission line, i.e., at $x = L$, can be estimated as a sum of the signals coming from each SQUID:
\begin{equation}
 V (t, L) = \sum_{k=0}^N \int \frac{d\omega}{2 \pi}  v(\omega) e^{- i \omega [t-(L-x_k)/c]}
 \label{eq:app_V}
\end{equation}
where $c$ is the speed of electromagnetic waves propagating along the transmission line. For simplicity, we assumed that the speed is the same both in the transmission line and in the chain of SQUIDs. Performing summation over $k$ in Eq. (\ref{eq:app_V}) and considering that $N \gg 1$, one finds
\begin{eqnarray}
 V (t, L) &=& \int \frac{d\omega}{2 \pi}  v(\omega) \frac{1- e^{- i  N \omega a/c}}{1- e^{- i  \omega a/c}}  e^{- i  \omega (t-L/c)} \nonumber \\
 &=& \sum_{n} \frac{V_n}{2 \pi}  \frac{1- e^{- i  N \omega a/c}}{1- e^{- i  \omega a/c}} e^{- i n \omega_0 (t-L/c)}.
\end{eqnarray}
Thus, the intensity of $n$-th harmonics at $x = L$ 
\begin{equation}
 P_n(L) = |V_n(L)|^2 = |V_n|^2 \Big| \frac{1- e^{- i  N \omega a/c}}{1- e^{- i  \omega a/c}}  \Big|^2 = P_n(0) \frac{\sin^2 \left( N n \frac{\omega_0 a}{2 c} \right)}{\sin^2 \left(  n \frac{\omega_0 a}{2 c}\right)}.
\end{equation}
The power of the $n$-th harmonics scales with the number of junction as $N^2$ if $ N a n \omega/ (2 c) \leq 1$.
This limit can be rewritten as $L/(2 \lambda) \leq 1$ with $N a =L$ and $c/(n \omega_0) =  \lambda_0/n = \lambda$ is the wavelength emitted.

For the parameters used in the simulation, considering a wavelength of $\lambda=6 \times 10^{-4}~$m corresponding to $200~$GHz in Silicon, and a SQUID distance of $a=1~\mu$m, we have that for $N=50$ the array can be considered as a lumped element.

\begin{figure}[t!]
\includegraphics[width=\columnwidth]{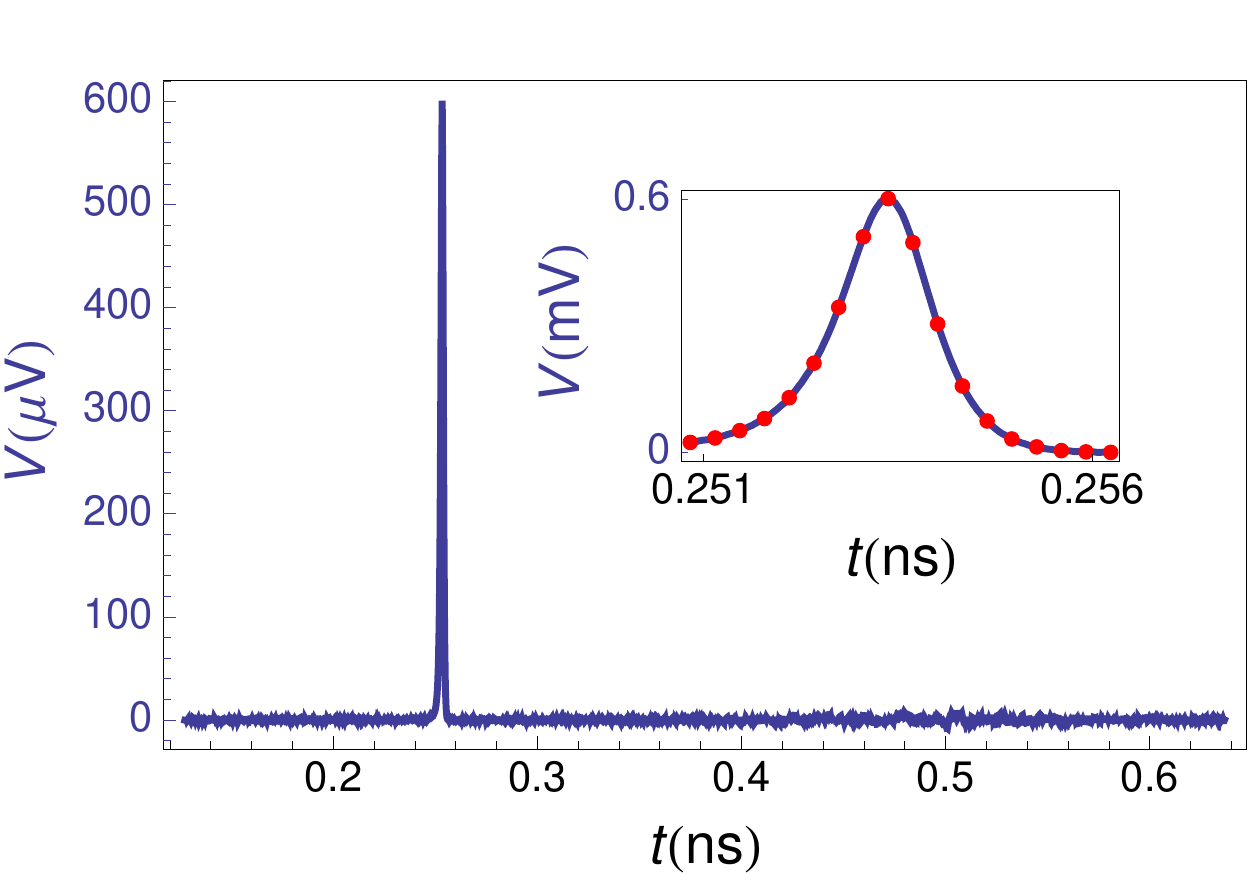}
\vspace{-6mm}
\caption{{\bf  Effect of the thermal noise on the dynamics.}
Voltage $V$ developed across the SQUID (blue solid line) as a function of time $t$ in presence of thermal noise. Here we set the driving frequency to $1~$GHz. 
(Inset) Comparison between noisy (blue solid curve) and noise-free dynamics (red dots).
The parameters are those typical of a YBCO junction, as reported in Refs. \cite{malnou2014high,rosenthal1994}: $I_+ R = 10$ mV and $I_B= 10^{-3}I_+$ and $\epsilon=0.9$. The capacity is negligible and set to zero. The temperature of the thermal bath is $4.2~$K.
 }
\label{fig1_SI}
\end{figure}

\section{Power delivered to the load resistor}

As discussed in the main text, current conservation through the SQUID implies that dynamics of each SQUID is independent from the rest of the chain. As a consequence, the voltage at the extremes of the chain scales as the number $N$ of SQUIDs. Accordingly, the intrinsic power generated by the device (that is, the power delivered to an infinite load) scales as $N^2$. 
In this section we discuss the effect of a finite load on the JRCG.
Let the SQUIDs be identical and $V(t)$ be the voltage drop across each SQUID. Then the voltage at the load is $V_{\rm tot}(t)=N V(t)$ and the current flowing through it is $I_L = N V(t)/R_L$, where $R_L$ is the (real) impedance of the load.
This current flows back to the SQUID array and adds up, with opposite sign, to the bias current $I_B$. As a result, \eqref{eq_app:RCSJ} must be rewritten as
\begin{equation}
  \frac{\hbar C}{2 e }  \ddot{{\varphi}} + \frac{\hbar}{2 e R} \dot{\varphi} + I_+  f(\varphi,t) = I_B - \frac{NV(t)}{R_L} \ .
\end{equation}
Recalling that $V_i(t)=(\hbar/2e)\dot\varphi$, we obtain
\begin{equation}
  \frac{\hbar C}{2 e }  \ddot{{\varphi}} + \frac{\hbar}{2 e \Reff} \dot{\varphi} + I_+  f(\varphi,t) = I_B \ ,
\end{equation}
with
\begin{equation}
 \Reff = \left( \frac{1}{R} +\frac{N}{R_L} \right)^{-1} = \frac{R R_L}{R_L + N R}\ . \label{eq:Reff}
\end{equation}

The effective change in the shunt resistance modifies the SQUID dynamics. In particular, it also reduces the power $P = N^2V^2/R_L$ that can be delivered to the load.
From Eq.~(\ref{eq:app_V^2}), we see that for a single SQUID $\langle V \rangle^2 \propto \Reff$.
Using \eqref{eq:Reff}, we find that $P$ scales as $N^2$ for $N \ll R_L/R$ and as $N$ for $N \gg R_L/R$.

This analysis suggests that an increase in the output power could be gained by operating more JRCGs in a parallel configuration. If $M$ such devices are operated in parallel and are synchronized, the backflow current is reduced by a factor $1/M$ and the effective resistance reads $\Reff = R R_L/ \left[R_L + (N/M)R \right]$.

\begin{figure}[t!]
\includegraphics[width=\columnwidth]{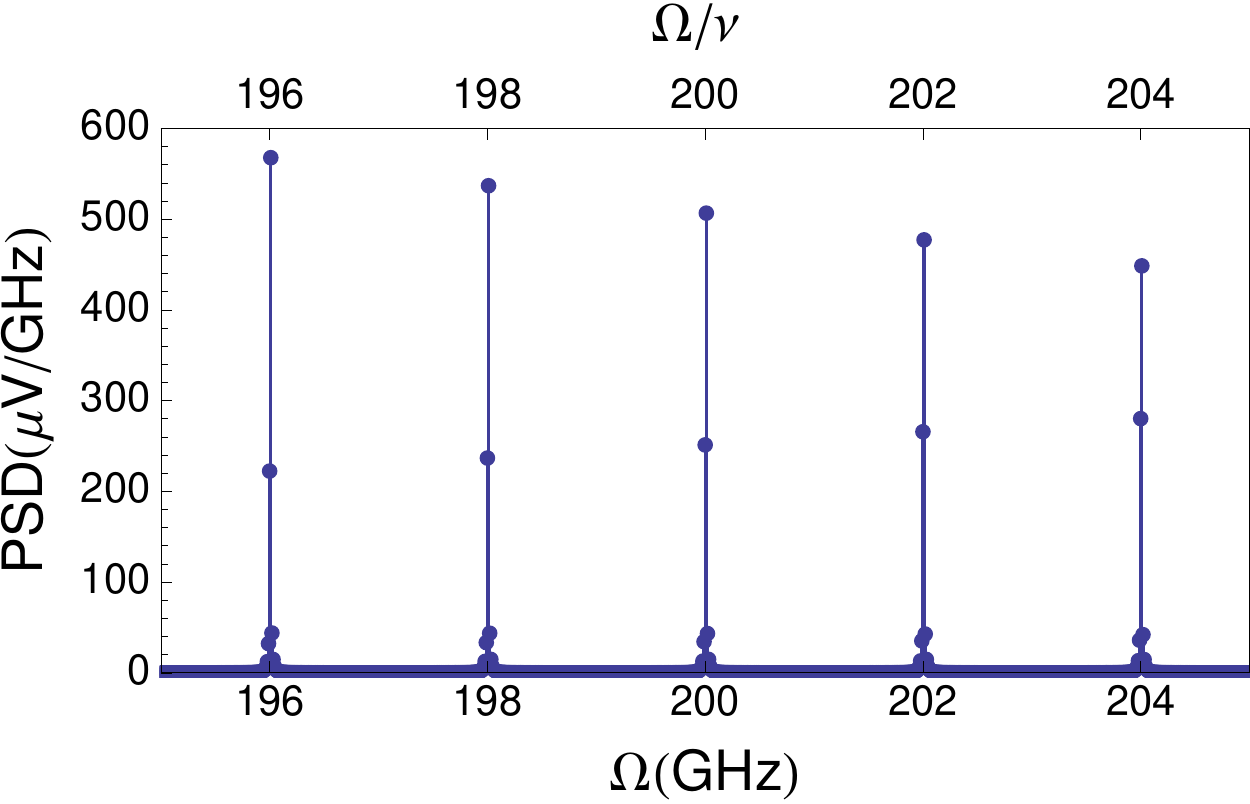}
\vspace{-6mm}
\caption{{\bf  Effect of the thermal noise on power spectral density.}
Power Spectral Density (PSD) for a YBCO SQUID with $1~$GHz drive at high frequency. The signal to noise ratio is about $10^3$. Notice that here the PSD is calculated for a single SQUID. For the calculations we set the following parameters \cite{malnou2014high,rosenthal1994}: $I_+ R = 10$ mV and $I_B= 10^{-3}I_+$. The junction capacitance is typically very small, $C\sim$ fF, and has been neglected in the numerics.
}
\label{fig2_SI}
\end{figure}


\section{Effect of the thermal noise}

To estimate the effect of the thermal noise we use the Langevin equation
\begin{equation}
	\frac{\hbar C}{2 e }  \ddot{{\varphi}} + \frac{\hbar}{2 e R} \dot{\varphi}- I_+ f(t) \sin \varphi = I_B + \xi(t),
	\label{eq:Langevin}
\end{equation}
where $ \xi(t)$ is the white noise with correlation function
\begin{equation}
 \langle \xi(t) \xi(t') \rangle = \frac{2 k_B T}{R} \delta(t-t').
\end{equation} 

We have numerically solved the associated stochastic differential equation in case of a YBCO SQUID with $1~$GHz drive.
We have considered a symmetric YBCO SQUID with negligible capacitance. The noise source has been taken at temperature of $4.2~$ K.
With these parameters, the dynamics of the SQUID shown in Fig. \ref{fig1_SI} is essentially identical to that without noise source.

The thermal noise results in a small broadening of the resonances at high frequency (see Fig. \ref{fig2_SI}).
The signal to noise ratio is still about $10^3$ and it can be further increased by decreasing the working temperature of the device.

The effect of noise is maximum when the energy barrier is shallow and undesired transitions are most likely to occur.
Therefore, we expect an increased noise influence for slow frequency drive since the system remains in a shallow barrier potential for a longer time.
However, even in this situation these noise effects can be reduced by increasing  $I_B$ in order to restore the privileged direction of the dynamics.

\section{Voltage spectrum and power}

To obtain the voltage power spectrum we first calculate the Fourier transform of the voltage $V(t)$
\begin{equation}
 V(\Omega) = \int_{0}^T dt e^{i \Omega t} V(t).
\end{equation}
The power spectral density (PSD) is then
\begin{equation}
  \text{PSD}(\Omega) = \frac{1}{T} | V(\Omega)|^2.
\end{equation}
The power $P$ discussed in the main text is calculated by integrating the PSD around the resonances $k \nu$ (where $\nu$ is the monochromatic drive frequency) and dividing for a standard load resistance of $50$~Ohm. This is the power we would measure {\it at a given resonance frequency} with a bandwidth exceeding the linewidth of the resonance.




\newpage
\begin{center}
  {\bf References}
\end{center}

\begin{thebibliography}{30}
\expandafter\ifx\csname natexlab\endcsname\relax\def\natexlab#1{#1}\fi
\expandafter\ifx\csname url\endcsname\relax
  \def\url#1{\texttt{#1}}\fi
\expandafter\ifx\csname urlprefix\endcsname\relax\def\urlprefix{URL }\fi

\bibitem[{Udem {\it et~al.}(2002)Udem, Holzwarth \&
  H{\"a}nsch}]{udem2002optical}
Udem, T., Holzwarth, R. \& H{\"a}nsch, T.~W.
\newblock Optical frequency metrology.
\newblock {\it Nature} \textbf{416}, 233--237 (2002).

\bibitem[{H{\"a}nsch \& Walther(1999)}]{hansch1999laser}
H{\"a}nsch, T. \& Walther, H.
\newblock Laser spectroscopy and quantum optics.
\newblock {\it Rev. Mod. Phys.} \textbf{71}, S242 (1999).

\bibitem[{Bloembergen(1977)}]{bloembergen1977nonlinear}
Bloembergen, N.
\newblock {\it Nonlinear spectroscopy}, vol.~64 (North Holland, 1977).

\bibitem[{H{\"a}nsch \& Inguscio(1994)}]{hansch1994frontiers}
H{\"a}nsch, T.~W. \& Inguscio, M.
\newblock {\it Frontiers in Laser Spectroscopy: Varenna on Lake Como, Villa
  Monastero, 23 June-3 July 1992}, vol. 120 (North Holland, 1994).

\bibitem[{Foreman {\it et~al.}(2007)Foreman, Holman, Hudson, Jones \&
  Ye}]{foreman2007remote}
Foreman, S.~M., Holman, K.~W., Hudson, D.~D., Jones, D.~J. \& Ye, J.
\newblock Remote transfer of ultrastable frequency references via fiber
  networks.
\newblock {\it Rev. Sci. Instrum.} \textbf{78}, 021101 (2007).

\bibitem[{Tinkham(2012)}]{tinkham2012introduction}
Tinkham, M.
\newblock {\it Introduction to superconductivity} (Courier Dover Publications,
  2012).

\bibitem[{Giazotto \& Mart{\'\i}nez-P{\'e}rez(2012)}]{giazotto2012josephson}
Giazotto, F. \& Mart{\'\i}nez-P{\'e}rez, M.~J.
\newblock The Josephson heat interferometer.
\newblock {\it Nature} \textbf{492}, 401--405 (2012).

\bibitem[{Giazotto {\it et~al.}(2013)Giazotto, Mart{\'\i}nez-P{\'e}rez \&
  Solinas}]{giazotto2013coherent}
Giazotto, F., Mart{\'\i}nez-P{\'e}rez, M. \& Solinas, P.
\newblock Coherent diffraction of thermal currents in Josephson tunnel
  junctions.
\newblock {\it Phys. Rev. B} \textbf{88}, 094506 (2013).

\bibitem[{Mart{\'\i}nez-P{\'e}rez \& Giazotto(2014)}]{martinez2014quantum}
Mart{\'\i}nez-P{\'e}rez, M.~J. \& Giazotto, F.
\newblock A quantum diffractor for thermal flux.
\newblock {\it Nat. Commun.} \textbf{5}, 3579 (2014).

\bibitem[{Patel \& Lukens(1999)}]{patel1999self}
Patel, V. \& Lukens, J.
\newblock Self-shunted Nb/AlO/sub x//Nb Josephson junctions.
\newblock {\it IEEE Trans Appl Supercond} \textbf{9},
  3247--3250 (1999).

\bibitem[{Gross \& Marx(2005)}]{gross2005applied}
Gross, R. \& Marx, A.
\newblock Applied superconductivity: Josephson effect and superconducting electronics.
\newblock {\it Walther-Mei{\ss}ner-Institut, Walther-Mei{\ss}ner-Str}
  \textbf{8}, 85748 (2005).

\bibitem[{Arutyunov {\it et~al.}(2008)Arutyunov, Golubev \&
  Zaikin}]{arutyunov2008superconductivity}
Arutyunov, K.~Y., Golubev, D.~S. \& Zaikin, A.~D.
\newblock Superconductivity in one dimension.
\newblock {\it Phys. Rep.} \textbf{464}, 1--70 (2008).

\bibitem[{Astafiev {\it et~al.}(2012)}]{astafiev2012coherent}
Astafiev, O. {\it et~al.}
\newblock Coherent quantum phase slip.
\newblock {\it Nature} \textbf{484}, 355--358 (2012).

\bibitem[{Langer \& Ambegaokar(1967)}]{langer1967intrinsic}
Langer, J.~S. \& Ambegaokar, V.
\newblock Intrinsic resistive transition in narrow superconducting channels.
\newblock {\it Phys. Rev.} \textbf{164}, 498 (1967).

\bibitem[{Zaikin {\it et~al.}(1997)Zaikin, Golubev, van Otterlo \&
  Zimanyi}]{zaikin1997quantum}
Zaikin, A.~D., Golubev, D.~S., van Otterlo, A. \& Zimanyi, G.~T.
\newblock Quantum phase slips and transport in ultrathin superconducting wires.
\newblock {\it Phys. Rev. Lett.} \textbf{78}, 1552 (1997).

\bibitem[{Likharev {\it et~al.}(1985)Likharev, Mukhanov \&
  Semenov}]{likharev1985resistive}
Likharev, K., Mukhanov, O. \& Semenov, V.
\newblock Resistive single flux quantum logic for the Josephson-junction
  digital technology.
\newblock {\it SQUID'85 Superconducting Quantum Interference Devices} (ed. Hahlbohm, H.-D. \& L\"ubbig, H.) 1103--1108 (de Gruyter, Berlin, 1985) ISBN: 978-3-11-086239-3.

\bibitem[{Mukhanov {\it et~al.}(1987)Mukhanov, Semenov \&
  Likharev}]{mukhanov1987ultimate}
Mukhanov, O., Semenov, V. \& Likharev, K.
\newblock Ultimate performance of the RSFQ logic circuits.
\newblock {\it IEEE Trans. Magn.} \textbf{23}, 759--762 (1987).

\bibitem[{Blais {\it et~al.}(2004)Blais, Huang, Wallraff, Girvin \&
  Schoelkopf}]{blais2004cavity}
Blais, A., Huang, R.-S., Wallraff, A., Girvin, S. \& Schoelkopf, R.~J.
\newblock Cavity quantum electrodynamics for superconducting electrical
  circuits: An architecture for quantum computation.
\newblock {\it Phys. Rev. A} \textbf{69}, 062320 (2004).

\bibitem[{Wallraff {\it et~al.}(2004)}]{wallraff2004strong}
Wallraff, A. {\it et~al.}
\newblock Strong coupling of a single photon to a superconducting qubit using
  circuit quantum electrodynamics.
\newblock {\it Nature} \textbf{431}, 162--167 (2004).

\bibitem[{Koch {\it et~al.}(2007)}]{koch2007charge}
Koch, J. {\it et~al.}
\newblock Charge-insensitive qubit design derived from the Cooper pair box.
\newblock {\it Phys. Rev. A} \textbf{76}, 042319 (2007).

\bibitem[{Malnou {\it et~al.}(2014)}]{malnou2014high}
Malnou, M. {\it et~al.}
\newblock High-Tc superconducting Josephson mixers for terahertz heterodyne
  detection.
\newblock {\it J. Appl. Phys.} \textbf{116}, 074505 (2014).

\bibitem[{Rosenthal \& Grossman(1994)}]{rosenthal1994}
Rosenthal, P. \& Grossman, E.~N.
\newblock Terahertz Shapiro steps in high temperature SNS Josephson junctions.
\newblock {\it IEEE Trans. Microw. Theory Tech.}
  \textbf{42}, 707--714 (1994).

\bibitem[{Shapiro(1963)}]{shapiro1963josephson}
Shapiro, S.
\newblock Josephson currents in superconducting tunneling: The effect of
  microwaves and other observations.
\newblock {\it Phys. Rev. Lett.} \textbf{11}, 80 (1963).

\bibitem[{Kautz \& Lloyd(1987)}]{kautz1987precision}
Kautz, R. \& Lloyd, F.~L.
\newblock Precision of series-array Josephson voltage standards.
\newblock {\it Appl. Phys. Lett.} \textbf{51}, 2043--2045 (1987).

\bibitem[{Tsai {\it et~al.}(1983)Tsai, Jain \& Lukens}]{tsai1983high}
Tsai, J.-S., Jain, A. \& Lukens, J.
\newblock High-precision test of the universality of the Josephson
  voltage-frequency relation.
\newblock {\it Phys. Rev. Lett.} \textbf{51}, 316 (1983).

\bibitem[{Lloyd {\it et~al.}({1987})}]{lloyd1987}
Lloyd, F.~L. {\it et~al.}
\newblock A Josephson array voltage standard at 10 V.
\newblock {\it IEEE Electron Device Lett.} \textbf{{8}}, {449--450}
  ({1987}).

\bibitem[{P{\"o}pel {\it et~al.}(1990)P{\"o}pel, Niemeyer, Fromknecht, Meier
  \& Grimm}]{popel19901}
P{\"o}pel, R., Niemeyer, J., Fromknecht, R., Meier, W. \& Grimm, L.
\newblock 1-and 10-V series array Josephson voltage standards in Nb/Al2O3/Nb
  technology.
\newblock {\it J. Appl. Phys.} \textbf{68}, 4294--4303 (1990).

\bibitem[{Cybart {\it et~al.}(2014)}]{cybart2014large}
Cybart, S.~A. {\it et~al.}
\newblock Large voltage modulation in magnetic field sensors from
  two-dimensional arrays of Y-Ba-Cu-O nano Josephson junctions.
\newblock {\it Appl. Phys. Lett.} \textbf{104}, 062601 (2014).

\bibitem[{Vicarelli {\it et~al.}(2012)}]{vicarelli2012graphene}
Vicarelli, L. {\it et~al.}
\newblock Graphene field-effect transistors as room-temperature terahertz
  detectors.
\newblock {\it Nat. Mater.} \textbf{11}, 865--871 (2012).

\bibitem[{Erickson {\it et~al.}(2014)Erickson, Vissers, Sandberg, Jefferts \&
  Pappas}]{erickson2014frequency}
Erickson, R., Vissers, M., Sandberg, M., Jefferts, S. \& Pappas, D.
\newblock Frequency Comb Generation in Superconducting Resonators.
\newblock {\it Phys. Rev. Lett.} \textbf{113}, 187002 (2014).

\end{thebibliography}


\end{document}